\documentstyle[12pt]{article}
\begin{document}

\newcommand{\beq}{\begin{equation}}
\newcommand{\eeq}{\end{equation}}

\begin{center} 
{\large{\bf Quantum theory of dissipation of a harmonic 
oscillator coupled to a nonequilibrium bath; Wigner-Weisskopf decay \\
and physical spectra}}

\vspace{1cm}
{\bf 
Jyotipratim Ray Chaudhuri, Bimalendu Deb, Gautam Gangopadhyay\footnote{
S. N. Bose National Centre for Basic Sciences, JD Block, Sector-III, 
Salt Lake City, Calcutta-700091, INDIA}
and\\ 
Deb Shankar Ray\\
Indian Association for the Cultivation of Science\\
Jadavpur, Calcutta-700032\\
INDIA }
\end{center}

\vspace{1cm}

\begin{abstract}
We extend the quantum theory of dissipation in the context of 
system-reservoir model, where the reservoir in question is kept in a
nonequilibrium condition. Based on a systematic separation of time scales
involved in the dynamics, appropriate generalizations of the fluctuation- 
dissipation and Einstein's relations have been pointed out. We show that
the Wigner-Weisskopf decay of the system mode results in a rate constant
which depending on the relaxation of nonequilibrium bath is dynamically modified. 
We also calculate the time-dependent spectra
of a cavity mode with a suitable gain when the cavity is kept in contact
with a nonequilibrium bath.
\end{abstract}

\vspace{2.5cm}


\newpage

\begin{center} 
{\bf I. INTRODUCTION}
\end{center} 

\vspace{0.5cm}  
     
     The problem of dissipative dynamics in quantum system is a key issue in 
     physics and chemistry today. The system-reservoir model describing the
     evolution of the quantum system coupled to a much larger system regarded
     as a reservoir has been the standard paradigm for dissipation in classical and quantum
     systems for many years [1-3].
     A popular variant of this model is the spin-boson
     model, a magnetic dipole coupled to a Boson field; another is a
     two-level atom in contact with a continuum of radiation field modes.
     These models describe a large variety of physical situations, such as,
     spontaneous emission, polaron formations, exiton motion, macroscopic quantum 
     tunneling etc., in atomic physics, solid state physics and quantum optics.

\vspace{0.2cm}
     
     In general, two very distinct situations emerge depending on the strength
     of the coupling constant between the system and the reservoir. 
     In the weak
     coupling case, the behavior of the former is only slightly affected by
     the reservoir which essentially behaves as a free field. In 
     the strong coupling
     cases (as in polaron theories)
     the polarization of the reservoir field
     by the system can not be ignored. The second important approximation 
     that is almost always made is that the correlation time of the reservoir
     must be very short enough (Markov approximation)
     for the interaction between 
     the system and the bath to be small (weak coupling approximation).
     Although
     several generalizations of the theory [3] which go beyond these 
     approximation 
     schemes are now available which describe the various interesting physical
     situations in condensed matter and quantum optical physics, one essential
     step is the assumption of an equilibrium distribution of the reservoir modes.
     Very little attention has been paid to the problem where the reservoir           
     in contact with the system is itself not in equilibrium.
     This nonstationarity
     of the bath is known to affect the kinetics of the system [4-6] leading to 
     nonexponential 
     decay in contrast to exponential decay in equilibrium  activation rate
     theories.
     Thus the relaxation of the nonequilibrium modes may 
     influence the dissipation of the system in question in a nontrivial way.
     The present paper addresses a related issue pertaining to
     quantum optical situations.

\vspace{0.2cm}
     
     We extend the quantum theory of dissipation of a harmonic oscillator
     coupled to a bath where the bath in question is in a nonequilibrum 
     condition.
     The nonequilibrium bath is effectively
     realized in terms of a semi-infinite dimensional broad-band reservoir
     which itself kept in contact with a thermal reservoir.
     We make use of
     a systematic time scale separation to construct the appropriate
     Langevin dynamics of the system mode. The fluctuation-dissipation
     and the Einstein relations have been suitably generalized.
     A detailed study of two model
     cases as immediate applications has been carried out.
     We show that
     the Wigner-Weisskopf decay rate constant of the system mode is dynamically
     modified when the system is coupled to a set of relaxing modes. We
     also calculate the transient noise-spectra of the cavity mode with a
     positive gain.
     The result is remarkably different from the steady-state
     spectra.

\vspace{0.2cm} 
     
     The outline of the paper is as follows: In section II we first  generalize
     the quantum theory of dissipation of a harmonic oscillator for a bath
      which is not in thermal equilibrium, followed by a derivation 
      of fluctuation-dissipation
      theorem in the next section III.
      Section IV is devoted to the discussion of  
      the Wigner-Weisskopf decay of the system.
      In section V we calculate the
      transient noise spectra of the system mode with a positive gain. The
      paper is concluded in section VI.

\vspace{1.0cm}

\begin{center}
{\bf II.  THE MODEL AND THE EQUATIONS OF MOTION}
\end{center}

\vspace{0.5cm} 
    
    To start with we consider a model consisting of a harmonic oscillator
    (the system) coupled to a set of relaxing modes considered as a semi-infinite
    dimensional system which effectively constitutes a nonequilibrium 
    reservoir. This in turn is in contact with a thermally equilibrated reservoir.
    Both the reservoirs are composed of two sets of harmonic oscillators characterized
    by the frequency sets \{$\omega_{j}$\} and \{$\Omega_{j}$\} for the
    equilibrium and nonequilibrium bath, respectively. The system-reservoir 
    combination develops in time under the influence of the total Hamiltonian
\newpage
\begin{eqnarray}
H & = & \hbar \omega_0 a^\dagger a + \hbar \sum_j \omega_j b_j^\dagger b_j
+\hbar \sum_\mu \Omega_\mu C_\mu^\dagger C_\mu \nonumber \\
& + & \hbar \sum_\mu g_\mu (C_\mu a^\dagger + C_\mu^\dagger a) +
\hbar \sum_\mu \sum_j \alpha_{j \mu} (b_j^\dagger C_\mu + b_j C_\mu^\dagger)\hspace{0.2cm}.
\end{eqnarray}

    The first term on the right-hand side describes the system mode 
    with characteristic frequency $\omega_0$ . The
    second and the third term represent the thermal and the nonequilibrium 
    linear modes. The next two terms represent the coupling of the nonequilibrium
    bath with the system mode and the thermal bath where the coupling constants
    are $g_{\mu}$ and $\alpha_{j \mu}$, respectively. In writing down the 
    Hamiltonian we have made use of the rotating wave approximation.

\vspace{0.2cm}
  
  The Heisenberg equations of motion for the system and the reservoir operators at any
  given time is given by
\begin{equation}
\dot{a}(t)=-i\omega_{0}a(t)-i \sum_{\mu} g_{\mu}C_{\mu}(t)\hspace{0.2cm},
\end{equation}
\begin{equation}
\dot{b}_{j}(t)=-i\omega_{j}b_{j}(t)-i \sum_{\mu} \alpha_{j\mu}C_{\mu}(t)\hspace{0.2cm},
\end{equation}
\begin{equation}  
\dot{C}_{\mu}(t)=-i\Omega_{\mu}C_{\mu}(t)-ig_{\mu}a(t)-i \sum_{j}\alpha_{j\mu}b_{j}(t)\hspace{0.2cm}.
\end{equation}   

Making use of the formal integral of the Eq.(3) for $b_{j}(t)$,
\begin{eqnarray*}  
b_{j}(t)=b_{j}(t_{0}) e^{-i\omega_{j}(t-t_{0})} -i\sum_{\mu}\alpha_{j\mu}
\int_{t_{0}}^{t}dt'C_{\mu}(t')e^{-i\omega_{j}(t-t')}
\end{eqnarray*}   

\noindent
in Eq.(4) we obtain
\begin{equation}
\dot{C}_{\mu}(t)=-i\Omega_{\mu}C_{\mu}(t)-ig_{\mu}a(t)-i\sum_{j}\alpha_{j\mu}
e^{-i\omega_{j}(t-t_{0})}b_{j}(t_{0})-\sum_{j}\sum_{\nu}\alpha_{j\mu}\alpha_{j\nu}
\int_{t{0}}^{t}dt'C_{\nu}(t')e^{-i\omega_{j}(t-t')}.
\end{equation}

Taking into consideration [1] that the interference time of
$\sum_{j}\alpha_{j\mu}\alpha_{j\nu}e^{-i\omega_{j}(t-t')}$ is much smaller
than the time over which the significant phase and amplitude modulation
of the linear modes $C_{\mu}(t)$ take place, the last term in Eq.(5)
can be identified as a relaxation term in the usual way with damping constant 
\begin{equation}
\gamma_{\mu\nu}^{c} = \pi \; \alpha_{\nu\mu}(\Omega_\nu) \;
\alpha_{\nu\nu}(\Omega_\nu) \; D(\Omega_\nu),
\end{equation}

\noindent
where $D(\Omega_\nu)$ represents the density of states of the 
equilibrium modes evaluated at 
$\Omega_\nu$.
Thus one can write down the Langevin equation of motion for the nonequilibrium
mode $C_\mu$ as follows;
\begin{equation}
\dot{C}_{\mu}(t)=-i\Omega_{\mu}C_{\mu}(t)-ig_{\mu}a(t) -
\sum_\nu\gamma_{\mu\nu}^cC_\nu(t)+f_{\mu}(t).
\end{equation}

Here the last term $f_{\mu}(t)$ represents the usual noise operator arising out of
the coupling of the nonequilibrium modes with the thermal bath modes as 
given by
\begin{equation}
f_{\mu}(t)=-i\sum_{j}\alpha_{j\mu}e^{-i\omega_{j}(t-t_{0})}b_{j}(t_{0})\hspace{0.2cm},
\end{equation}

\noindent
with reservoir average of $f_{\mu}(t)$ is zero, i.e.,
\begin{equation}
{\langle f_{\mu}(t) \rangle}_{B}=0\hspace{0.2cm},
\end{equation}

\noindent
where by the average ${\langle O(t) \rangle}_{B}$ of an operator $O(t)$ we
mean ${\langle O(t) \rangle}_{B} = Tr \{ O(t) \rho_{B} \}$. Here
$\rho_{B}$ denotes the initial density operator for the thermal bath $\{b_{j}\}$
and is a multimode extention of the usual thermal operator. This is given by 
\begin{eqnarray*}
\rho_{B} = \prod_{j} [exp \{- \frac {(\omega_{j} {b_{j}^{\dagger}} b_{j})}
{KT}\}][1 - exp (  \frac {\omega_{j}} {KT} )] \hspace{0.2cm}, 
\end{eqnarray*}

\noindent
where $T$ is the equilibrium temperature. Note that in defining the average
we assumed the initial factorization of the total density operator into
subsystem densities for the system, thermal bath and the nonequilibrium bath.

Introducing the slowly varying operator
\begin{equation}
\tilde{C}_{\mu}(t)=C_{\mu}(t)e^{i\Omega_{\mu}(t-t_{0})}
\end{equation}

\noindent
the Eq.(7) reduces to the following form;
\begin{equation}
\dot{\tilde{C}}_{\mu}(t)=-ig_{\mu}a(t)e^{i\Omega_{\mu}(t-t_{0})} -
\sum_{\nu}\gamma_{\mu\nu}^c \tilde{C}_{\nu}(t)e^{i(\Omega_{\mu}-\Omega_{\nu})(t-
t_{0})}+F_{\mu}(t)\hspace{0.2cm},
\end{equation}

\noindent
where
\begin{equation}
F_{\mu}(t)=f_{\mu}(t)e^{i\Omega_{\mu}(t-t_{0})}\hspace{0.2cm}.
\end{equation}

The relevant properties of the noise operator $F_{\mu}(t)$ 
can be summarized as;
\begin{equation}
{\langle F_{\mu}(t) \rangle}_{B}=0\hspace{0.2cm},
\end{equation}

\noindent
and
\begin{equation}
{\langle F_{\mu}^{\dagger}(t)F_{\nu}(t') \rangle}_{B} =
\gamma_{\mu\nu}^c \bar{N}(\Omega_{\mu})
\delta(t-t')\delta_{\mu\nu}\hspace{0.2cm}.
\end{equation}

The last relation follows from 
\begin{equation}
{\langle b_{m}^{\dagger}(t_{0})b_{n}(t_{0})\rangle}_{B}
=\bar{N}(\omega_{n})\delta_{mn}\hspace{0.2cm}, 
\end{equation}

\noindent
where $\bar{N}(\omega_{n})$ is the thermal average of the number 
operator of the equilibrium bath and is given by 
$\bar{N}(\omega_{n}) = \frac{1}{exp(\frac{\omega_n}{KT}) - 1}$ .
Also note that $\delta_{\mu\nu}$ takes care secular approximation. Eq.(14)
also implies a purely Ohmic frequency-independent dissipation of the 
nonequilibrium modes.

\vspace{0.2cm}

Taking into consideration of the standard fluctuation-dissipation 
relation for the thermal 
bath in terms of Eq.(14), we arrive at the following Langevin equation for
the nonequilibrium bath modes;
\begin{equation}
\dot{C}_{\mu}(t)=-i\Omega_{\mu}C_{\mu}(t) - ig_{\mu}a(t) - \gamma_{\mu\mu}^{c}
C_{\mu}(t)+f_{\mu}(t)\hspace{0.2cm}.
\end{equation}

\vspace{0.2cm}

Eq.(16) constitutes an important result of this section which takes into account
of the relaxation of the intermediate oscillator modes due to their coupling
to the standard thermal bath whose fluctuations are described by $f_{\mu}(t)$.
It is important to note that the above consideration is based on the rotating
wave approximations (RWA) which is frequently used when considering coupling
to a heat bath in quantum optics. In the present case as we are to see in the
subsequent sections that there are quite subtle interaction effects. The question
whether these effects would survive a more complete treatment may naturally
arise. We mention two pertinent points at this stage. First, if one retains
the nonrotating couplings in the Hamiltonian and carry out the same perturbative
procedure one arrives at an equation of motion for $C_{\mu}(t)$ [instead of
Eq.(11) ]  which additionally contains secular oscillating terms of the form
$C_{\nu}^{\dagger} e^{i(\Omega_{\mu}+\Omega_{\nu})(t-t_{0})}$ . RWA amounts
to neglecting these contributions which may be important only at very high 
coupling. Secondly, C$_{\mu}$ modes executes a slow relaxation dynamics on
the time scale $\sim 1/\gamma_{\mu\mu}^{c}$ as compared to the time scale of 
correlation of thermal noise. The time scale of secular oscillations being  
short, those can be safely averaged out from the relevant dynamics.

\vspace{0.2cm}
Another important point regarding RWA in the context of the present linear
coupling model Hamiltonian (1) needs to be considered. Ford, O'connell and
Lewis [13] have demonstrated that independent oscillator model within
RWA (a variant of LC model) is seriously flawed since the Hamiltonian becomes
imaginary when the bath is not passive ( i.e., there exists an associated
spectrum of eigenvalues ranging upto $-\infty$ ). The problem essentially
lies at the specific frequency dependence of the coupling constant in its 
denominator (for example, as shown in Ref. [13], appropriately transformed
$g_{j}$ in Eq. (1) is 
proportional to $\frac{1}{\sqrt{\omega_0 \omega_j}}$ ). However, a standard
procedure in quantum optics is to replace the summation over modes by an 
integral over their density $D(\omega_j)$, (in free space equal to
$(\frac{V \omega_{k}^{2}}{c^{3} \pi^{2}})$), i.e.,
$\sum_{j} (g_{j})^{2} \rightarrow \int_{-\infty}^{+\infty} d\omega  D(\omega) 
g^{2}(\omega)$ \hspace{0.2cm}. One thus gets rid of the unwanted frequency
dependence in the denominator of the coupling constant in the calculations
(see, for example, calculation of Wigner-Weisskopf decay rate, Lamb et al in
Ref.[1] ). Since we have followed the same approach, RWA does not pose
any problem in the present analysis. The problem however, remains for a 
strictly discrete spectrum.

\vspace{0.2cm}
We have presented above an extension of the quantum theory of damping from
Langevin point of view within a traditional system-reservoir linear coupling
scheme. Essentially the model consists of replacing the reservoir by
damping terms in the Heisenberg equations of motion for dissipation-free
system and adding fluctuating forces as driving terms which add fluctuations
to the system. The operator forces are such that (i) the system has the correct
statistical properties to agree in the classical limit and (ii) they maintain
the commutation relations for Boson operators to ensure that uncertainty
principle is not violated. These considerations have been fully taken care
of in our analysis with appropriate elaboration in the following sections. The 
spiritual root of the quantum statistical approach to damping lies in the 
fluctuation-dissipation relation, which illustrates a dynamical balance of 
inward flow of energy due to fluctuations from the reservoir into the 
system and the outward flow of energy from the system to the reservoir due
to the dissipation of the system mode. We address this specific issue in the
next section.

\vspace{0.5cm}

\begin{center}
{\bf III.  FLUCTUATION-DISSIPATION RELATION FOR NONEQUILIBRIUM BATH} 
\end{center}

\vspace{0.5cm}

To explore the influence of an initial excitation of the semi-infinite
dimensional intermediate reservoir modes and its relaxation, we now consider the
evolution of these linear modes $C_{\mu}$ in terms of Eq.(16). The physical
situation that has been addressed is the following;

\vspace{0.2cm}
We consider that at $t=t_{0}$ the
excitation is switched on and the bath modes $(C_{\mu},C_{\mu}^{\dagger})$
are thrown into a nonstationary
state such that they behave as a nonequilibrium reservoir undergoing
relaxation. We follow the stochastic dynamics of the system mode and the 
relaxation of the nonequilibrium reservoir modes after $t > t_{0}$. We
assume that the effect of back reaction of the system mode on the reservoir 
modes is small enough to be neglected. Eq.(16) allows a formal solution of 
the following form

\vspace{0.2cm}
\beq
C_{\mu}(t) = C_{\mu}^{s}(t) + C_{\mu}(t_{0}) e^{(-i\Omega_{\mu}-\gamma_{\mu\mu}    
^{c}) (t-t_{0})} - ig_{\mu}\int_{t_{0}}^{t}dt'e^{(-i\Omega_{\mu}-\gamma_{\mu\mu}  
^{c}) (t-t')} a(t') \hspace{0.2cm}.
\eeq

\vspace{0.2cm}

The first term on the right hand side in the absence of the coupling of the system
mode represents the ( long time ) stationary stochastic solution of the form

\vspace{0.2cm}
\begin{equation}
C_{\mu}^{s}(t) = C_{\mu}^{s} e^{-i[\Omega_{\mu}(t-t_{0}) +
\phi_{\mu}^{s}]} \hspace{0.2cm},
\end{equation}

\vspace{0.2cm}

\noindent
where the amplitude $C_{\mu}^{s}$(operators) and phases $\phi_{\mu}^{s}$
(c-numbers) are assumed to 
be randomly distributed. The random nature of $C_{\mu}^{s}(t)$ may be
understood in the following way : Let us first note that in the absence
of coupling $g_{\mu}$ Eq.(16) allows the solution

\vspace{0.2cm}
\begin{eqnarray*}
C_{\mu}(t) =  C_{\mu}(t_0) 
e^{-(i \Omega_{\mu} + \gamma_{\mu\mu}^{c} ) (t-t_0) } 
+ \; e^{-(i \Omega_{\mu} + \gamma_{\mu\mu}^{c} ) (t-t_0)}
\int_{t_0}^{t} dt' f_\mu(t')
\; e^{(i \Omega_{\mu} + \gamma_{\mu\mu}^{c}) (t'-t_0)} \; \; .
\end{eqnarray*}

\noindent

In the steady state we neglect the first term which decays rapidly. The 
second term is a randomly fluctuating term (which is the 
most important term in almost any Langevin analysis) 
due to $f_\mu(t)$. Substituting Eq.(8) in 
the above equation we obtain

\vspace{0.2cm}
\begin{eqnarray*}
C_{\mu}^{s}(t) = e^{-i \Omega_{\mu}(t-t_0)} \sum_{j} ( -i \alpha_{j \mu} ) \;
e^{-\gamma_{\mu\mu}^c (t-t_0)}  \; b_{j}(t_0)
\int_{t_0}^{t} dt' \; e^{(i\omega_{j} + i \Omega_{\mu} + 
\gamma_{\mu\mu}^{c} ) (t'-t_0)} \; \; ,
\end{eqnarray*}

\vspace{0.2cm}

\noindent
where $s$ signifies the steady state. The above solution implies that
$C_{\mu}^{s}(t)$ is essentially a superposition of unknown (since the
initial condition for the infinite number of thermal bath oscillator
$b_j(t_0)$ are assumed to be completely uncertain) amplitudes (operators)
and phases ($c$-numbers) and may be written compactly in the form of 
Eq.(18). When written in the form (18) we obtain an instantaneous
realization of the random distribution of $C_{\mu}^{s}$ and $\phi_{\mu}^s$.
Thus Eq.(17) represents an instantaneous solution of Eq.(16).
To check the consistency of the solution (17)
we note the following points; (i) To recover initial deterministic 
amplitude of $C_{\mu}(t)$ we need an ensemble average of Eq.(17) at 
$t=t_{0}$ so that ${\bar{C}}_{\mu}^{s}(t_{0})=0$. Also we are to see 
(subsequent to Eq.(24)) that the first term in Eq.(17) is responsible 
for the usual fluctuation-dissipation relation when the nonequilibrium 
modes get equilibrated at $t=\infty$ . $C_{\mu}^{s}(t)$ in $C_{\mu}(t)$ 
is thus dictated by the condition of stationarity. (iii) It can be easily seen
that the random nature 
of the operator forces in $C_{\mu}^{s}(t)$ is responsible for maintaining 
the Boson commutation relation for $C_{\mu}(t)$ which  further
ensures that the uncertainty principle is not violated.

\vspace{0.2cm}

The second term on the right hand side in Eq.(17) carries the information
of relaxation of the $C_{\mu}$ modes due to their coupling to the thermal bath
and is a typical memory term. The third term on the other hand represents
the effect of coupling of the system mode to the nonequilibrium reservoir.

\vspace{0.2cm}

Substitution of the solution(17) in Eq.(2) yields the equation of motion 
for the slowly varying system operator $A(t)$;

\vspace{0.2cm}
\begin{equation}  
\dot{A}(t)=-\sum_{\mu}g_{\mu}^{2}\int_{t_{0}}^{t}dt'e^{i(\omega_{0}-\Omega_{\mu})
(t-t')}e^{-\gamma_{\mu\mu}^{c}(t-t')}A(t') \; + \; Z(t),
\end{equation}

\vspace{0.2cm}

\noindent
where

\vspace{0.2cm}
\begin{equation}  
A(t)=a(t)e^{-i\omega_{0}(t-t_{0})}
\end{equation}

\vspace{0.2cm}

\noindent
and

\vspace{0.2cm}
\begin{equation}
Z(t)=-i\sum_{\mu}g_{\mu}[C_{\mu}^{s}(t)+C_{\mu}(t_{0})
e^{(-i\Omega_{\mu}-\gamma_{\mu\mu}^{c})(t-t_{0})}]e^{i\omega_{0}(t-t_{0})}.
\end{equation}

\vspace{0.2cm}

Eq.(19) is a non-Markovian equation where the memory effects arise out 
of the two sources; the first one being the system operator concerned term
$A(t')$ which depends on earlier time $t'$; the other one $e^{-\gamma_{\mu\mu}
^{c}(t-t')}$ is due to the relaxation of the nonequilibrium bath modes 
which arises because of their coupling to the thermal bath. $Z(t)$ represents
the noise operator for the nonequilibrium bath modes.

\vspace{0.2cm}

In the weak coupling approximation scheme the first term in equation(19)
can be simplified to the following form;

\vspace{0.2cm}
\begin{eqnarray*}
A(t)\int d\Omega\rho(\Omega)g^{2}(\Omega)\int_{0}^{\infty}d\tau 
e^{i(\omega_{0}-\Omega)\tau}e^{-\gamma_{\mu\mu}^{c}\tau},
\end{eqnarray*}

\vspace{0.2cm}

\noindent
where the summation over the bath modes is replaced by integration
and $\rho(\Omega)$ represents (an a priory known) density of nonequilibrium 
bath modes. Assuming weak dependence of $\gamma_{\mu\mu}^{c}$ on the modes 
one can reduce the last expression (after performing integration over 
$\Omega$) to obtain the following Langevin equation for the system operator;

\vspace{0.2cm}
\begin{equation}
\dot{A}(t)=-\Gamma \; A(t) + Z(t) ,
\end{equation}

\vspace{0.2cm}

\noindent
where

\vspace{0.2cm}
\begin{equation}
\Gamma = \pi g^{2}(\omega_{0})\rho(\omega_{0})
\end{equation}

\vspace{0.2cm}

\noindent
can be identified as a dissipation constant of the system mode due to the
fluctuations of the nonequilibrium reservoir in the limit when $\gamma_{\mu\mu}
^{c}$ is vanishingly small. The explicit dependence of the decay constant
of the system mode on $\gamma$ will be revealed clearly ( as a more general
case ) in the next section 
within the scope of Wigner-Weisskopf approximation.

\vspace{0.2cm}

The nonequilibrium generalization of the fluctuation-dissipation relation
is now immediately apparent. From the expression for $Z(t)$ (Eq.(21))
one finds that
\begin{eqnarray}
{\langle Z^{+}(t)Z(t') \rangle}_{NR} & = & \sum_{\mu}g_{\mu}^{2}
[{\langle C_{\mu}^{s \dagger}C_{\mu}^{s} \rangle}_{NR}
e^{i(\Omega_{\mu}-\omega_{0})(t-t')} \nonumber \\
& + & {\langle C_{\mu}^{\dagger}
(t_{0})C_{\mu}(t_{0}) \rangle}_{{\rm NR}}e^{i(\Omega_{\mu}-\omega_{0})(t-t')}
e^{2\gamma_{\mu\mu}^{c}t_{0}}e^{-\gamma_{\mu\mu}^{c}(t+t')}] \hspace{0.2cm}.
\end{eqnarray}

\vspace{0.2cm}

We denote the average photon number of the nonequilibrium bath by

\vspace{0.2cm}
\begin{equation}
\bar{n}(\Omega_{\mu},t_{0}) ={\langle C_{\mu}^{\dagger}(t_{0})C_{\mu}(t_{0}) 
\rangle}_{{\rm NR}}\hspace{0.2cm},
\end{equation}

\vspace{0.2cm}

\noindent
where $t_{0}$ signifies the dependence of average photon number of the 
nonequilibrium bath on its initial state of preparation. Also the steady 
state average photon number is given by 

\vspace{0.2cm}
\begin{eqnarray*}
\bar{n}(\Omega_{\mu}) =\langle C_{\mu}^{s \dagger} C_{\mu}^{s} 
\rangle_{{\rm NR}}\hspace{0.2cm}.
\end{eqnarray*}

\noindent
By $\langle O(t) \rangle_{NR}$ we mean  $\langle O(t) \rangle_{NR} =
Tr\{O(t) \rho_c\}$ where $\rho_c$ indicates the initial thermalized
density operator for the intermediate oscillator $\{C\}$-modes, and is given by
\begin{eqnarray*}
\rho_c = \prod_{\mu} exp(-\frac{\Omega_\mu C_\mu^\dagger C_\mu}{KT}) [1-
exp(\frac{\Omega_\mu}{KT})] \hspace{0.2cm}.
\end{eqnarray*}

\noindent
As usual, the initial factorization of densities of $\{b\}$ and $\{C\}$
modes is assumed.

\vspace{0.2cm}  

After replacing the summation by integration and $\gamma_{\mu\mu}^{c}$ by the 
average $\gamma$ in Eq.(24) we obtain

\vspace{0.2cm}  
\begin{eqnarray*}
\langle Z^{\dagger}(t)Z(t')\rangle_{{\rm NR}}=\left[ \Gamma
{\bar{n}} (\omega_{0})+e^{-2\gamma(t-t_{0})} \Gamma
{\bar{n}}(\omega_{0},t_{0})\right] \delta(t-t')\hspace{0.2cm}.
\end{eqnarray*}

\vspace{0.2cm}

Rewriting $\Gamma {\bar{n}}(\omega_{0},t_{0})$ in terms of a 
deviation from its steady state value $\Gamma {\bar{n}} (\omega_{0})$ as

\vspace{0.2cm}
\begin{eqnarray*}
\Gamma {\bar{n}}(\omega_{0},t_{0})=D(t_{0})-
\Gamma {\bar{n}}(\omega_{0}) \hspace{0.2cm},
\end{eqnarray*}

\vspace{0.2cm}

\noindent
we identify a time-dependent diffusion coefficient $D(t)$ in the last 
equation as 

\vspace{0.2cm}
\begin{eqnarray*}
D(t)=\Gamma {\bar{n}}(\omega_{0})+\left [ D(t_{0})-
\Gamma {\bar{n}}(\omega_{0})\right ] e^{-2\gamma(t-t_{0})}
\hspace{0.2cm}.
\end{eqnarray*}

\vspace{0.2cm}

We thus obtain

\vspace{0.2cm}
\beq
\langle Z^{\dagger}(t)Z(t')\rangle_{{\rm NR}}=\left\{
\Gamma {\bar{n}}(\omega_{0})+\left [ D(t_{0})-
\Gamma {\bar{n}}(\omega_{0})\right ] e^{-2\gamma(t-t_{0})}
\right \} \delta(t-t') \hspace{0.2cm}.
\eeq

\vspace{0.2cm}

Eq.(26) is the desired nonequilibrium generalization of the 
fluctuation-dissipation relationship. This relates instantaneous fluctuations of the
nonequilibrium bath (which itself is undergoing relaxation at a rate
$\gamma$ due to its coupling with the thermal bath) to the 
dissipation of the energy of the system mode through $\Gamma$. The 
nonequilibrium nature of the bath is implicit in the initial 
preparation which creates an initial diffusion coefficient $D(t_{0})$
and also in the exponentially decaying term. In the long time
limit one recovers the usual fluctuation-dissipation relation for the thermal
bath at equilibrium. 

\vspace{0.2cm}
We now express Eq.(26) in terms of energy density of fluctuations of the
nonequilibrium modes. The energy density which is proportional to the
power spectrum centered around $\omega_0$ is given by [$\hbar = 1$]
\begin{eqnarray*}
u(\Omega,t) & = & \frac{\Omega}{4 \Gamma} \int_{-\infty}^{+\infty} d\tau 
\langle Z^{\dagger}(t) Z(t+\tau) \rangle e^{i(\Omega-\omega_0)\tau} \\
& = & \frac{1}{2} \Omega \bar{n}(\Omega) + e^{-2 \gamma (t-t_0)} 
[ u(\Omega,t_0) - \frac{1}{2} \Omega \bar{n}(\Omega) ] \; \; .
\end{eqnarray*}

\vspace{0.2cm}

\noindent
It is important to note that $t$ is the slow time variable which is well
separated from the time scale of thermal noise. The fluctuations of the noise operator 
$Z(t)$ is now explicitly determined by the nonequilibrium state of the intermediate 
oscillator bath through its energy density $u(\Omega,t)$ at each instant
of time $t$. In other words the instantaneous nonequilibrium energy 
density distribution
of the fluctuating modes is related to the friction coefficient of these modes on the system
degree of freedom through a dynamic equilibrium. One can immediately recover 
the classical version of the above equation in the high temperature limit
( where $\bar{n}(\Omega) = \frac{1}{e^{\frac{\Omega}{KT}} - 1} \simeq \frac
{KT}{\Omega}$ ) to obtain

\vspace{0.2cm}
\begin{eqnarray*}
u(\Omega,t) = \frac{1}{2} KT + e^{-2 \gamma(t-t_0)} \left[u(\Omega,t_0)-
\frac{1}{2}KT \right] \; \; . 
\end{eqnarray*}

\vspace{0.2cm}

\noindent
The above equation was derived earlier [6] in the context of classical
relaxation kinetics of complex nonlinear systems. This reduction to classical version 
of the nonequilibrium fluctuation-dissipation relation serves as a consistency
check for its quantum generalization (26) which is more relevant in quantum
optical issues that we address in this paper.

\vspace{0.2cm}

Another point should be emphasized regarding the fluctuation-dissipation
relation (26). The very notion of a nonequilibrium bath apparently
suggests that frequency distribution function of the modes of this bath
should be a function of time and in principle, one should seek
for it as a self-consistent solution from a quantum kinetic analysis.
We have followed here an alternative route. Since we are working with
Heisenberg operator equations of motion, the knowledge of initial total
state density which is factorizable in subsystem densities (for the system,
thermal bath and nonequilibrium or intermediate bath ) at $t=0$, is sufficient
to describe the complete dynamics in terms of average values and correlation 
functions. We are concerned here (and also in classical theory [6]) with 
a priori given frequency distribution functions for equilibrium $\{b_j\}$ 
bath ($D(\omega)$) and nonequilibrium $\{C_\mu\}$ bath $(\rho(\Omega))$
which are independent of time (nor we have introduced any non-thermal
temperature in describing the nonequilibrium bath). The essential
content of the nonequilibrium nature of the bath rests on a time-dependent
energy density fluctuation distribution function $u(\Omega,t)$ as described 
above (varying over a slower time scale
compared to the time scale of thermal noise ) which is a derived
quantity rather than a self-consistently obtained function which might be
obtainable from a quantum kinetic analysis. The effect of initial excitation
is to create an energy density function $u(\Omega,0)$ which differs from
its equilibrium counterpart. This departure sets in a nonequilibrium
situation. All these considerations also apply to the classical version 
of nonequilibrium fluctuation-dissipation relation [6].

\vspace{0.2cm}

Before bringing an end to this section some pertinent points are to be noted.
First, we need to stress that in the derivation of the relation (26) we
assume that $Z(t)$ is effectively stationary on the fast correlation
time scale of the thermal bath. Second, the theory as developed above
is based on the consideration of quantum optical situations in mind. It is
well known [2,11] that, in general, the noise from the equilibrium bath at low
temperatures is very different from a simple white noise and concerns 
expressions which contain integral from distribution function of the bath
over all frequencies. Most often such situations are encountered in condensed
matter and in chemical physics of complex systems. However in the problems of 
quantum optics where the harmonic oscillator bath serves as a standard paradigm
for optical fields one routinely uses a broad band white noise spectrum
such that $\sum_{\mu}g_{\mu}^{2}{\bar{n_{\mu}}}$ is slowly varying and the 
summand in Eq.(24) is so strongly peaked at $\Omega_{\mu}=\omega_{0}$, that we may
convert the sum into an integral and remove the slowly varying factors to 
obtain the result (26) [ p. 422 of Louisell in Ref. 1 ]. Similar consideration leads
us to Einstein's spontaneous emission coefficient or Wigner-Weisskopf decay
rate which contains single frequency $\omega_{0}$, the characteristic
frequency of the system mode. Thus although, in principle, it may be possible
to consider a colored noise spectrum and the resulting frequency dependence
of rate constants we restrict ourselves to former situations of broad band
reservoir which itself is undergoing a relaxation on the time scale of $1/ \gamma$.
An important content of the present work is to explore the effect of this
secondary relaxation on the primary kinetics of the system mode. The 
nonequilibrium generalization of the fluctuation dissipation relation as discussed
in this section serves as a basis of this exploration in following two quantum
optical cases.

\vspace{0.5cm}

\begin{center}
{\bf IV. DECAY OF THE SYSTEM MODE ; WIGNER-WEISSKOPF APPROXIMATION}
\end{center}

\vspace{0.5cm}

We now obtain the solution of Heisenberg-Langevin integro-differential
equation of motion [Eq.(19) ] for the system mode coupled 
to nonequilibrium bath under Wigner-Weisskopf approximation. 
If we take  the Laplace
transformation of Eq.(19), we have after some algebra 
\begin{equation}
\bar{A}(s) =\frac{a(0)+\bar{Z}(s)}{s+\sum_{\mu} \frac{g_{\mu}^{2}}
{s+i(\Omega_{\mu}-\omega_{0})+ \gamma_{\mu\mu}^{c}}} ,
\end{equation}

\noindent
where
\begin{equation}
\bar{A}(s) = \int_{0}^{\infty}dtA(t)e^{-s t}
\end{equation}

\noindent
and
\begin{eqnarray*}
\bar{Z}(s)=\int_{0}^{\infty}dtZ(t)e^{-s t} ,
\end{eqnarray*}

\noindent
or,
\beq
\bar{Z}(s)=-i \sum_{\mu}g_{\mu}C_{\mu}(0)\frac{1}{s+i(\Omega_{\mu}-
\omega_{0})+\gamma_{\mu\mu}^{c}} \hspace{0.2cm},
\eeq

\vspace{0.2cm} 

\noindent
where we have used the fact that the amplitudes  and the phases of $C_{\mu}
^{s}(0)$ are random. Also we have $A(0)=a(0)$. Here we have chosen the initial 
time $t_{0}=0$ for convenience.

\vspace{0.2cm} 

We make Wigner-Weisskopf approximation to solve for the zeros of 
$\bigtriangleup$ in Eq.(27), where

\vspace{0.2cm} 
\beq
\frac{1}{\bigtriangleup} = (s+\sum_{\mu}\frac{g_{\mu}^{2}}{s+i(\Omega_{\mu}-
\omega_{0})+\gamma_{\mu\mu}^{c}})^{-1} .
\eeq

\vspace{0.2cm}

For weak interaction, zeroth approximation is $\bigtriangleup=0$ if
$s=0$. As a next approximation let $s \rightarrow 0$ in the denominator of 
the sum in $\bigtriangleup$. In other words under 
Wigner-Weisskopf approximation we calculate the first order shift in the 
simple pole which is approximately given by

\vspace{0.2cm} 
\beq 
\bigtriangleup(0) - s \simeq Lt_{s \rightarrow 0}
\sum_{\mu}\frac{g_{\mu}^{2}}{s+i(\Omega_{\mu}-
\omega_{0})+\gamma_{\mu\mu}^{c}}
= \gamma^W + i \delta\omega
\eeq

\vspace{0.2cm} 

\noindent
where $\gamma^W$ and $\delta\omega$ are real quantities.
Explicit calculation in the usual way yields,

\vspace{0.2cm} 
\beq
\gamma^W = \int d\Omega \rho (\Omega)g^2(\Omega) 
\frac{\gamma}{{(\Omega-\omega_0)}^2+\gamma^2},
\eeq

\vspace{0.2cm} 

\noindent
and

\vspace{0.2cm} 
\beq
\delta\omega = \int d\Omega \rho(\Omega)g^2(\Omega)
\frac{(\Omega-\omega_0)}{{(\Omega-\omega_0)}^2+\gamma^2}.
\eeq

\vspace{0.2cm}

The expressions for the line width $\gamma^W$ and the frequency
shift $\delta\omega$ of the system mode thus obtained due to the relaxation 
of the nonequilibrium bath are markedly different from the usual expressions 
of Wigner-Weisskopf theory. This is because of the explicit dependence of 
$\gamma^W$ and $\delta\omega$ on the $\gamma$ in Eqs.(32)
and (33) which arises due to the relaxation of the nonequilibrium modes
due to their coupling with the thermal bath. In the limit 
$\gamma\rightarrow 0$ 
one recovers the usual decay rate $\Gamma$ and the level-shift terms, i.e.,

\vspace{0.2cm}
\begin{eqnarray*}
\lim_{\gamma \rightarrow 0} \gamma^{W}=\Gamma  \hspace{0.2cm}.
\end{eqnarray*}

\vspace{0.2cm}

We thus see that the effect of Wigner-Weisskopf approximation is to
replace the more exact equation(19) by the Langevin equation whose 
solution obtained after appropriate inverse Laplace transform of equation(27)
is given by

\vspace{0.2cm} 
\begin{eqnarray}
A(t) & = & a(0)e^{-(\gamma^W+i \delta \omega)t}-
\sum_{\mu} g_{\mu}C_{\mu}(0)\frac{e^{-\gamma t}e^{-i (\Omega_{\mu}-\omega_{0})t}
\left[ 1-e^{-(\gamma^W-\gamma)t} e^{i(\Omega_{\mu}-\omega_{0}-
\delta\omega)t}\right ]}{(\omega_{0}-\Omega_{\mu}+\delta\omega)-i(
\gamma^W-\gamma)} \hspace{0.2cm}.
\end{eqnarray}

\vspace{0.2cm}

It may be noted that while deriving $\gamma^{W}$ and $\delta\omega$, weak
coupling and smooth density of states $\rho(\Omega)$ for the intermediate
oscillators have been assumed. By the same token, the correction terms in Eq.
(34) are small and the result for $A(t)$ as expressed in Eq.(34) could be 
simplified further to the following form,

\vspace{0.2cm}  
\beq
A(t)=a(0)e^{-(\gamma^W +i\delta\omega)t} +\chi(\omega_{0}) 
C_{\omega_{0}}(0)\left[ e^{-\gamma t}-e^{-(\gamma^W + i\delta
\omega)t}\right ]\hspace{0.2cm},
\eeq

\vspace{0.2cm}  

\noindent
where $\chi(\omega_{0})$ is given by

\vspace{0.2cm}   
\begin{eqnarray*}
\chi(\omega_{0})=\left[ \frac{g(\omega_{0}) \rho(\omega_{0})}
{\delta\omega-i\left(\gamma^W - \gamma\right)}\right]\hspace{0.2cm}.
\end{eqnarray*}

\vspace{0.2cm}   

\noindent
The primary evolution of the system mode $A$ thus depends on the secondary 
relaxation of the intermediate oscillators explicitly.

\vspace{0.2cm}   

The expression for $\gamma^W$ (Eq.(32)) illustrates a dynamical
modification of the Wigner-Weisskopf decay rate constant since it
incorporates the effect of coupling of the nonequilibrium bath to the
thermal bath through $\gamma$. The modification of atomic spontaneous 
decay rate both in the form of enhancement and suppression by
appropriate tailoring of vacuum modes of the cavity is now well known
in cavity QED [7]. Whereas in the cavity QED problems one essentially
77manipulates the boundary conditions in various ways, the present modification is 
effectively dynamical in nature in the sense that it carries the effect of
relaxation of the nonequilibrium modes on the dissipation of the
system mode. It is then also expected that the atomic decay rate might 
be similarly affected in appropriately modified situations.

\vspace{0.2cm}

Before completing this section we point out that in the present problem
of quantum theory of dissipation in the quantum optical context
we are concerned with the frequency spectrum of the radiation field modes.
In the context of solid state the frequency density is assumed to be
of the Debye type with appropriate regularization by cutoff at high
frequency. In the cavity QED problems adjustment of boundary conditions
may lead to different density of states. Although there is no generalization
of the dependence of friction on the frequency spectrum, in general,
one encounters a time-retarded friction.

\vspace{0.5cm}

\begin{center}
{\bf V.  TIME-DEPENDENT SPECTRUM OF A CAVITY MODE WITH GAIN IN CONTACT WITH
NONEQUILIBRIUM BATH}
\end{center}

\vspace{0.5cm}

An immediate consequence of the nonequilibrium generalization
of the fluctuation-dissipation relation is the explicit time dependence
of the diffusion constant, as evident in Eq.(26). It is therefore
expected that this time-dependence may make its presence felt if one
analyses the transient noise spectrum of the system mode. With this
end in view we now calculate the time-dependent spectrum of a cavity 
field mode coupled to a nonequilibrium reservoir which causes the field
mode to decay at the rate $\Gamma$. Generally we
find the spectrum by applying the quantum regression theorem to the Langevin
equation for a quantized field mode interacting with a medium described 
by a gain $\alpha$. In general, the complex gain $\alpha(t)$ is an operator
that is saturated by the number operator $A^{\dagger}(t) A(t)$ [p. 467 of
Ref.12]. The Langevin equation for our problem is given by,
\begin{equation}
\dot{A}(t)=-(\Gamma+i \delta-\alpha)A(t)+Z(t) \hspace{0.2cm},
\end{equation}

\vspace{0.2cm}

\noindent
where $A(t)$ and $Z(t)$ are slowly varying annihilation and noise operators,
respectively. In general, Eq.(36) applies to laser-like situations 
including those with two-level and semiconductor media [12].
Here $\delta(=\omega_0-\nu)$ is the detuning of the mode oscillation
frequency $\nu$ from the passive cavity resonance frequency $\omega_0$
and $\alpha$, the gain coefficient is assumed to be a real number
$(\Gamma > \alpha)$. $Z(t)$ is the noise source operator as given by

\vspace{0.2cm}
\beq
Z(t)=-i \sum_\mu g_\mu [C_\mu^s(t)+C_\mu(t_0)
e^{(-i \Omega_\mu-i \nu - \gamma_{\mu \mu}^C)(t-t_0)}].
\eeq

The noise is characterized by the following properties :

\vspace{0.2cm}
\beq
{\langle Z(t) \rangle}_{NR}=0,
\eeq

\vspace{0.2cm}
\beq
{\langle Z^{\dagger}(t) Z(t') \rangle}_{NR}=\left[ \Gamma {\bar{n}}+
\{D(t_{0})-\Gamma {\bar{n}}\} e^{-2\gamma(t-t_{0})}\right] \delta(t-t')
\hspace{0.2cm}.
\eeq

\vspace{0.2cm}

It has also to be noted that since we are dealing with a non-stationary
situation the standard steady state spectrum is not applicable. We therefore 
take resort to non-steady state spectrum or the so-called
``physical spectrum'' of the cavity mode [8] where the attention is focused 
on the dynamic evolution of the spectrum following an abrupt excitation of
a near-resonant cavity mode. The main reason
for studying the time-dependent spectrum is that the familiar power spectrum of the
Wiener-Khintchine theorem is not applicable to nonstationary processes. Although
in quite a number of earlier cases time-dependent spectrum of Page and
Lampard were used widely serious objections were raised against this spectrum
(e.g., it can be negative). Eberly and Wodkiewicz have shown that the suitably
normalized counting rate of a photodetector can be used to define a time-dependent
spectrum or physical spectrum. This allows the influence of the spectrum
analyzer (basically a Fabry-Perot interferometer, for example) to be exhibited
in the spectrum so that the band limit of the measuring device is appropriately
incorporated which makes the spectrum free from ambiguities and unphysical
characteristics of earlier spectrum. It has also been emphasized [9]
that when the instrument width, $W$ is narrow enough such that $W \ll \Gamma$
, the spectrum appears to be qualitatively similar to Wiener-Khintchine 
spectrum. This transient spectrum
has been used earlier in several occasions in connection with 
resonance fluorescence studies [9], micromaser problem [10] etc. 
One can define the time-dependent spectrum or the physical spectrum as 
follows;

\vspace{0.2cm}
\beq
S(t, \omega, W)=2 W Re \int_0^t dt_2 e^{-W(t-t_2)}
\int_0^{t-t_2} d\tau e^{(\frac{W}{2}-i \Delta)\tau}
\langle A^\dagger (t_2+\tau) A(t_2) \rangle.
\eeq

\vspace{0.2cm}

Here the symbols have the following meaning : $t$ is the elapsed time after
the system and the reservoir have been subjected to initial excitation at
$t=t_{0}(=0), W$
is the full width of the transmission peak of the interferometer and
$\Delta(=\omega-\nu)$ is the detuning, or frequency offset of the Fabry-Perot
line center above the frequency of the field $\omega$. It is important to note 
that the time-dependent spectrum is expressed in terms of two integrals in
Eq.(40). The first integral is over the correlation time and is actually
the counterpart of the Wiener-Khintchine spectrum band limited by the width of 
the measuring device, $W$, while the second one over $t_{2}$ takes into account
of the nonstationarity which makes the correlation function $t_{2}$ 
dependent. The device width in the second integral also sets the limit over
the time scale of this nonstationarity.

\vspace{0.2cm}

Since the time-evolution of the system is governed by Eq.(36), one can make 
use of the quantum regression hypothesis which yields two-time correlation
function

\vspace{0.2cm}
\beq
\langle A^\dagger (t+\tau) A(t) \rangle =
e^{-(\Gamma-i \delta-\alpha)\tau}
\langle A^\dagger (t) A(t) \rangle.
\eeq

\vspace{0.2cm}

We emphasize that $t$ in Eq.(41) [or $t_2$ in Eq.(40)] refers to the
non-stationary time. We therefore calculate the explicit time
dependence of $\langle A^\dagger (t) A(t)\rangle$ using Einstein's relations
(see Appendix A for details);

\vspace{0.2cm}
\beq
\frac{d}{dt} \langle A^\dagger (t)A(t) \rangle =
-2(\Gamma-\alpha) \langle A^\dagger (t) A(t) \rangle +
2 \Gamma \bar{n}[1+r e^{-2 \gamma t}]\hspace{0.2cm},\hspace{0.2cm}
{\rm where}\hspace{0.2cm}r=\frac{D(0)}{D(\infty)}-1 \hspace{0.2cm}.
\eeq

\vspace{0.2cm}

The solution of Eq.(42) is given by

\vspace{0.2cm}
\begin{eqnarray}
\langle A^{\dagger}(t)A(t) \rangle  
& = & e^{-2(\Gamma-\alpha)t} \langle A^{\dagger}(0)A(0) \rangle
\hspace{9.0cm}\nonumber\\
\nonumber\\
& & + \Gamma \bar{n}(\omega_{0}) \left[
\frac{1}{\Gamma-\alpha}-
\frac{(1+r)(\Gamma-\alpha)-\gamma}{(\Gamma-\alpha)(\Gamma-\alpha-\gamma)}
e^{-2(\Gamma-\alpha)t}+r
\frac{e^{-2 \gamma t}}{(\Gamma-\alpha-\gamma)}\right] .
\end{eqnarray}

\vspace{0.2cm}   

From Eq.(42) we obtain the steady state condition

\vspace{0.2cm}   
\beq
(\Gamma-\alpha)\langle A^\dagger (\infty)A(\infty) \rangle
= \Gamma \bar{n} .
\eeq

\vspace{0.2cm}   

We let $\langle A^\dagger (\infty)A(\infty) \rangle =N$. Eqs.(44) and
(43) may then be rewritten, respectively, as

\vspace{0.2cm}   
\beq
\bar{n}=\frac{\Gamma-\alpha}{\Gamma} N
\eeq

\vspace{0.2cm}   

\noindent
and

\vspace{0.2cm}
\beq
\langle A^\dagger(t)A(t) \rangle = N(1-rke^{-2 a t}+r k e^{-2 \gamma t}),
\eeq

\vspace{0.2cm}   

\noindent
where $a=\Gamma-\alpha$, {\large $k=\frac{a}{a-\gamma}$}\hspace {0.2cm}.

\vspace{0.2cm}   

Combining Eqs.(41) and (46) we obtain the two-time correlation
function

\vspace{0.2cm}   
\beq
\langle A^\dagger(t_2+\tau)A(t_2) \rangle 
= e^{-(\Gamma-i \delta-\alpha)\tau}
N(1-r k e^{-2 a t_2}+r k e^{-2 \gamma t_2}).
\eeq

\vspace{0.2cm}   

Making use of this relation and performing the integration over
$\tau$ and $t_2$, we extract the real part [Eq.(40)] which yields

\newpage   
\begin{eqnarray}
S(t, \Delta, W) & = & \frac{2 N W}{W_{-}^{2}+\Delta^{2}}\left[ \frac{W_{-}}{W_{+}}
\left\{(1+k r )e^{-2Wt}-\frac{1}{2} k r e^{-2(W+a)t}-(1+\frac{1}{2}k r)
e^{-Wt}\right\}\right.\nonumber\\
\nonumber\\
& & \frac{W_{+}W_{-}+\Delta^{2}}{W_{+}^{2}+\Delta^{2}}\left\{1+k r e^{-2\gamma t}
\right\}+k r e^{-2at} - \frac{(1+kr)2a\Delta}{W_{+}^{2}+\Delta^{2}}
e^{W_{+}t}\sin \Delta t \nonumber\\
\nonumber\\
& & \left.+\frac{2akr-(W_{+}W_{-}+\Delta^{2})}{W_{+}^{2}+\Delta^{2}} e^{-W_{+}t}
\cos \Delta t\right]\hspace{0.2cm},
\end{eqnarray}

\vspace{0.2cm}

\noindent
where, $W_{+} =\frac{W}{2}+(\Gamma-\alpha) \hspace{0.2cm} {\rm and } \hspace{0.2cm}
W_{-}=\frac{W}{2}-(\Gamma-\alpha) \hspace{0.2cm}$.

\vspace{0.2cm}

Here we have set the detuning $\delta=0$. 

\vspace{0.2cm}

In the long limit $(t \rightarrow\infty)$ the spectrum reaches the steady 
state value

\vspace{0.2cm}
\begin{equation}
S(\Delta, W)=\frac{2 N W}{{\left\{\frac{W}{2}-(\Gamma-\alpha)\right\}}^2
+\Delta^2}
\left[\frac{
{\left(\frac{W}{2}\right)}^2-{(\Gamma-\alpha)}^2+\Delta^2}
{
{\left\{\frac{W}{2}+(\Gamma-\alpha)\right\}}^2+\Delta^2} \right] .
\end{equation}

\vspace{0.2cm}

It is interesting to note that at short time the effect of 
nonequilibrium bath is prominent through $\gamma$ and $r$. 
While $r(=\frac{D(0)}{D(\infty)} -1)$ includes the effect of preparation of 
the initial nonequilibrium condition by a sudden external excitation at $t=0$
which makes the initial diffusion coefficient $D(0)$ to be different from its
stationary long time value $D(\infty)$, $\gamma$ carries the information
of relaxation. As expected, the steady state spectrum is independent of
both $\gamma$ and $r$.
This is because at large time when the nonequilibrium bath is
equilibrated, the system forgets its past and the time-dependence of 
diffusion coefficient is erased and
the spectrum becomes the steady state spectrum.

\vspace{0.2cm}

We now look for the transient characteristics of the spectra of the cavity
mode. In Fig 1 we plot the spectra for different scaled time $t$ after
the initial excitation, $\Gamma$ being used as a scaling parameter.
For numerical computation we choose the following scaled parameter set :
$W = 4, \alpha = 0.1, r=1$ and $\gamma = 0.1$. One observes that after
the initial excitation the spectra grow to a maximum and then the peak
height start diminishing and eventually reaches the steady state value.
Thus the effect of relaxation of the nonequilibrium bath becomes prominent
in the short time region. The variation of peak intensity with time
for the three different $\gamma$ has been shown in Fig2. It is apparent
that the spectrum reaches the steady state more quickly for larger values
of $\gamma$ and also for small $\gamma$ the maximum peak is larger
than that for larger $\gamma$.

\vspace{0.5cm}

\begin{center}
{\bf VI. CONCLUSIONS}
\end{center}

\vspace{0.5cm}

We have developed the quantum theory of dissipation of a harmonic
oscillator coupled to a nonequilibrium bath in terms of a
microscopic model. Making use of appropriate separation of time
scales one can construct an effective Langevin dynamics with memory
(where the memory functions are not the phenomenological inputs)
which is due to the relaxation of the nonequilibrium bath modes
and identify the relevant noise sources. An essential offshoot is the
nonequilibrium generalizations of the familiar fluctuation-dissipation 
and Einstein's relations. It is important to note that the Wigner-Weisskopf decay rate
constant of the oscillator is dynamically modified. The theory is further
applied to calculate the time-dependent spectrum of a cavity mode with
suitable gain. One observes that the nonequilibrium nature of the bath 
modes makes its presence felt in the time-dependence of the diffusion
constant leading to the differential behavior of the transient spectra
from the steady-state ones. Although in the present problem we are mainly
concerned with the dissipation of energy, it may also be worthwhile to
investigate the problem of decoherence on similar footing. We hope to 
address this and related issues in future.

\vspace{1cm}

{\bf Acknowledgment :} Thanks are due to the Department of Science and
Technology (Govt.of India) for partial financial support.

\newpage

\begin{center}
{\bf Appendix A}\\
\underline{Generalized Einstein's relations}
\end{center}

\vspace{0.5cm}

In this section we outline the derivation of the nonequilibrium 
generalization of Einstein's relations Eq.(41).

\vspace{0.2cm}

The system operators follow the Langevin equation of motion Eq.(35)

\vspace{0.2cm}
\begin{equation}
\dot{A} = -(\Gamma+i \delta-\alpha)A+Z(t) \; \; ; 
\end{equation}

\vspace{0.2cm}

\noindent
where the first term within the parenthesis of the right hand
side is the drift term and $Z(t)$ is the nonequilibrium noise operator
[Eq.(37)] . From the identity

\vspace{0.2cm}
\begin{equation}
{A}^\dagger(t)=A^\dagger(t-\Delta t)+\int_{t-\Delta t}^t
dt' \dot{A}^{\dagger} (t')
\end{equation}

\vspace{0.2cm}

\noindent
we first obtain the correlation function of the system and
the noise operator ;

\vspace{0.2cm}
\begin{equation}
\langle A^\dagger(t)Z(t) \rangle =
\langle A^\dagger(t-\Delta t)Z(t) \rangle+
\int_{t-\Delta t}^t dt' \langle [-(\Gamma-i \delta-\alpha)A^\dagger(t')
+Z(t')]Z(t) \rangle .
\end{equation}

\vspace{0.2cm}

Because the operator $A^\dagger(t')$ at time $t'$ is not affected
by fluctuation at a later time $t$, the first term on the right hand
side is zero. Similarly the correlation $\langle A^\dagger(t')Z(t) \rangle$
is zero except at the point $t'=t$ but the integration is zero. 
Thus we have

\vspace{0.2cm}
\begin{equation}
\langle A^\dagger(t)Z(t) \rangle = \int_{t-\Delta t}^{t}dt'
\langle Z^\dagger(t')Z(t) \rangle.
\end{equation}

\vspace{0.2cm}

Note that we have not assumed the stationary property
of the noise.

\vspace{0.2cm}

Next we determine the equation of motion for the average
$\langle A^\dagger(t)A(t) \rangle$;

\vspace{0.2cm}
\begin{equation}
\frac{d}{dt} \langle A^\dagger(t)A(t) \rangle =
\langle \dot{A}^\dagger(t)A(t) \rangle +
\langle A^\dagger(t)\dot{A}(t) \rangle.
\end{equation}

\vspace{0.2cm}

From Eq.(50) we have after some algebra

\vspace{0.2cm}
\begin{equation}
\frac{d}{dt} \langle A^\dagger(t)A(t) \rangle =
-2(\Gamma-\alpha)\langle A^\dagger(t)A(t) \rangle +
\langle Z^\dagger(t)A(t) \rangle +
\langle A^\dagger(t)Z(t) \rangle.
\end{equation}

\vspace{0.2cm}

Substituting Eq.(52) and its adjoint in Eq.(55) and performing the
integral over $t'$ where we use Eq.(39), we obtain the nonequilibrium
generalization of Einstein's relation

\vspace{0.2cm}
\begin{equation}
\frac{d}{dt} \langle A^\dagger(t)A(t) \rangle =
-2(\Gamma-\alpha)\langle A^\dagger(t)A(t) \rangle +
2 \Gamma \bar{n}(\omega_{0})\left[ 1+\left\{\frac{D(t_{0})}{D(\infty)}-1\right
\} e^{-2\gamma(t-t_{0})}\right]\hspace{0.2cm},
\end{equation}

\vspace{0.2cm}

\noindent
where we have used the notation $D(\infty)=\Gamma \bar{n}(\omega_{0})$.

\newpage

{\bf References:}

\begin{enumerate}
\item  W. H. Louisell,  Quantum Statistical Properties of Radiation (Wiley,
 New York 1973) ;
 M. Lax,  Phys. Rev.  {\bf 145}, 110 (1966) ;
 M. Lax and H.Yuen,  Phys. Rev. {\bf 172}, 362 (1968) ;
 G. S. Agarwal,  Phys. Rev.  {\bf A 2}, 2038 (1970) ;
 R. Graham and H. Haken,  Z. Phys. {\bf 235},  166 (1970) ;
 M. Sergent III, M. O. Schully, W. E. Lamb, Jr., Laser Physics (Addison-Wesley,
 Massachusetts 1974).
\item A. Caldeira and A. J. Leggett,  Ann. Phys. {\bf 149}, 374 (1983) ;
\item See for example, a recent review,
G. Gangopadhyay and D. S. Ray, in Advances in Multiphoton Processes and
Spectroscopy, edited by S. H. Lin, A. A. Villayes and 
F. Fujimura (World Scientific,
Singapore, 1993) vol. 8 ;
 G. Gangopadhyay and D. S. Ray ,  Phys. Rev. {\bf A 46}, 1507 (1992); 
 Phys. Rev.
{\bf A 43}, 6424 (1991) ; J. Chem. Phys. {\bf 96}  4693 (1992);
 C. W. Gardiner and M. J. Collect ,   
 Phys. Rev. {\bf A 31},  3761 (1985);
 A. E. Ekert and P. L. Knight,     Phys. Rev.  {\bf A 47}. 487 (1990) .
\item R. Landauer,    J. Stat. Phys.  {\bf 9},  351 (1973); {\bf 11},  525  (1974);
{\bf 13},  1  (1975).
\item D. L. Stein, R. Doering, R. G. Palmer, J. L. van Hemmen and R. M. McLaughlin ,  
 Phys. Letts. {\bf A 136},  353 (1989) .
\item M. Millonas and C. Ray,   Phys. Rev. Letts.  {\bf 75}, 1110 (1995).
\item See, for example, 
S. Haroche and D. Kleppner,    Phys. Today {\bf 42}, 24 (1989);
E. A. Hinds, in Advances in Atomic,
Molecular and Optical Physics {\bf 28},
237 (1991); 
D. Meschede,   Phys. Reps. {\bf 219}, 263 (1992);
Cavity Quantum Electrodynamics, edited by P. R. Berman (Academic Press, 1994) .
\item J. H. Eberly and Wodkiewicz ,    
J. Opt. Soc. Am.  {\bf 67},  1252  (1977) .
\item J. H. Eberly , C. V. Kunasz and K. Wodkiewicz ,   
J. Phys. {\bf B 13}, 217 (1980) .
\item B. Deb and D. S. Ray , Phys. Rev. {\bf A 49},  5015 (1994).
\item A. Schmid, J. Low. Temp. Phys.   {\bf 49}, 609 (1982).
\item P. Meystre and M. Sargent III, Elements of Quantum Optics (Springer-
Verlag, Berlin 1990).
\item G. W. Ford, R. F. O'Connell and J. T. Lewis, 
Phys. Rev. {\bf A 37},  4419 (1988).
\end{enumerate}

\newpage

\begin{center}
{\bf Figure Captions}
\end{center}

1. Time-dependent spectra of the cavity mode with gain for different
dimensionless times \\ 
(a) $\Gamma t = 0.05$, (b) $\Gamma t = 0.5$,
(c) $\Gamma t = 3.0$, (d) $\Gamma t = 10.0$ with $\alpha=0.1$, 
$\delta=0.0$, $W=4.0$ and $r=1.0$ (Scales arbitrary).

\vspace{2cm}

2. Variation of intensity of the time-dependent spectra vs.
dimensionless time for different dimensioless decay rate constants
of the nonequilibrium modes; \\
(a) $\frac{\gamma}{\Gamma} = 0.1$,
(b) $\frac{\gamma}{\Gamma} = 0.2$, (c) $\frac{\gamma}{\Gamma} = 0.3$
with $\alpha, \delta, W$ and $r$ same as Fig. 1
(Scales arbitrary).
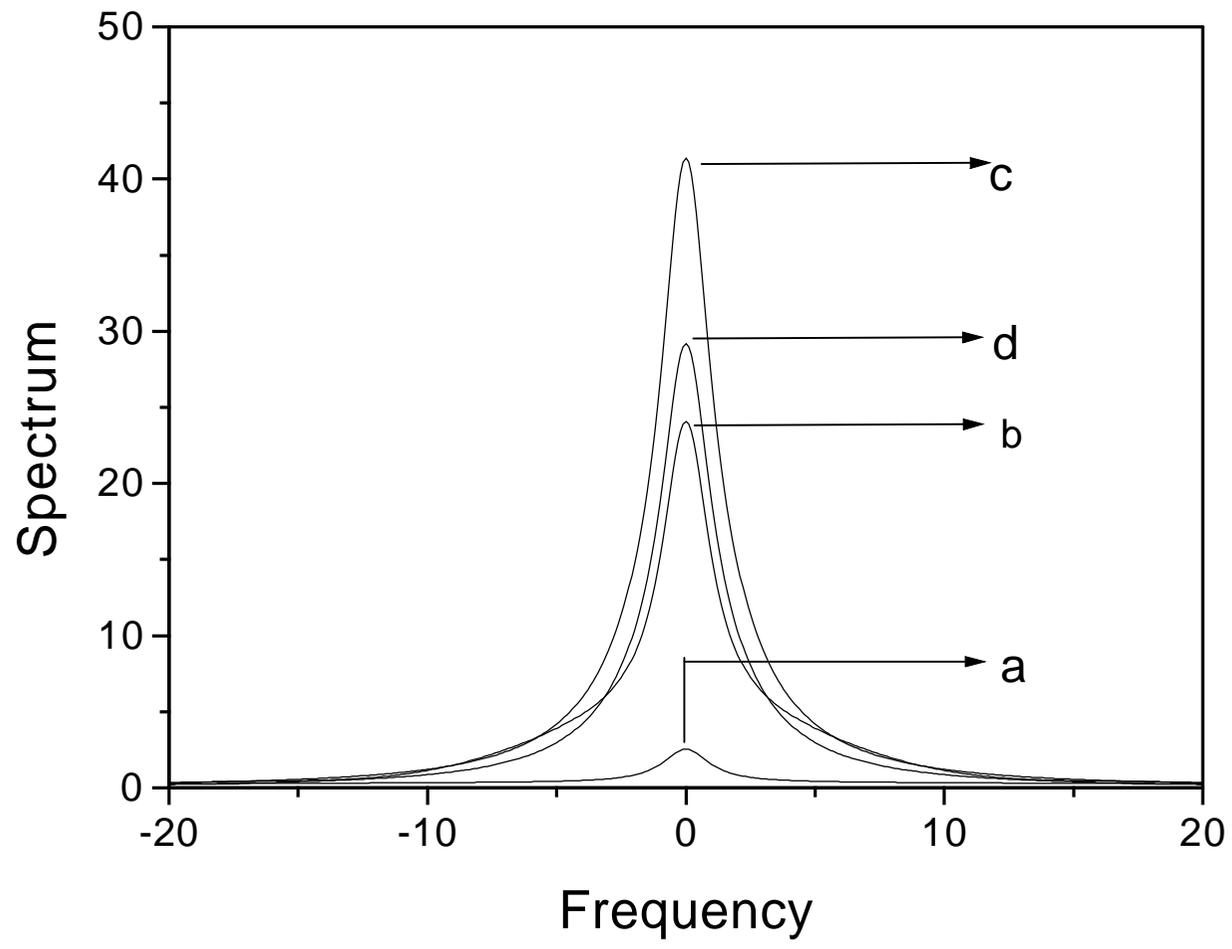
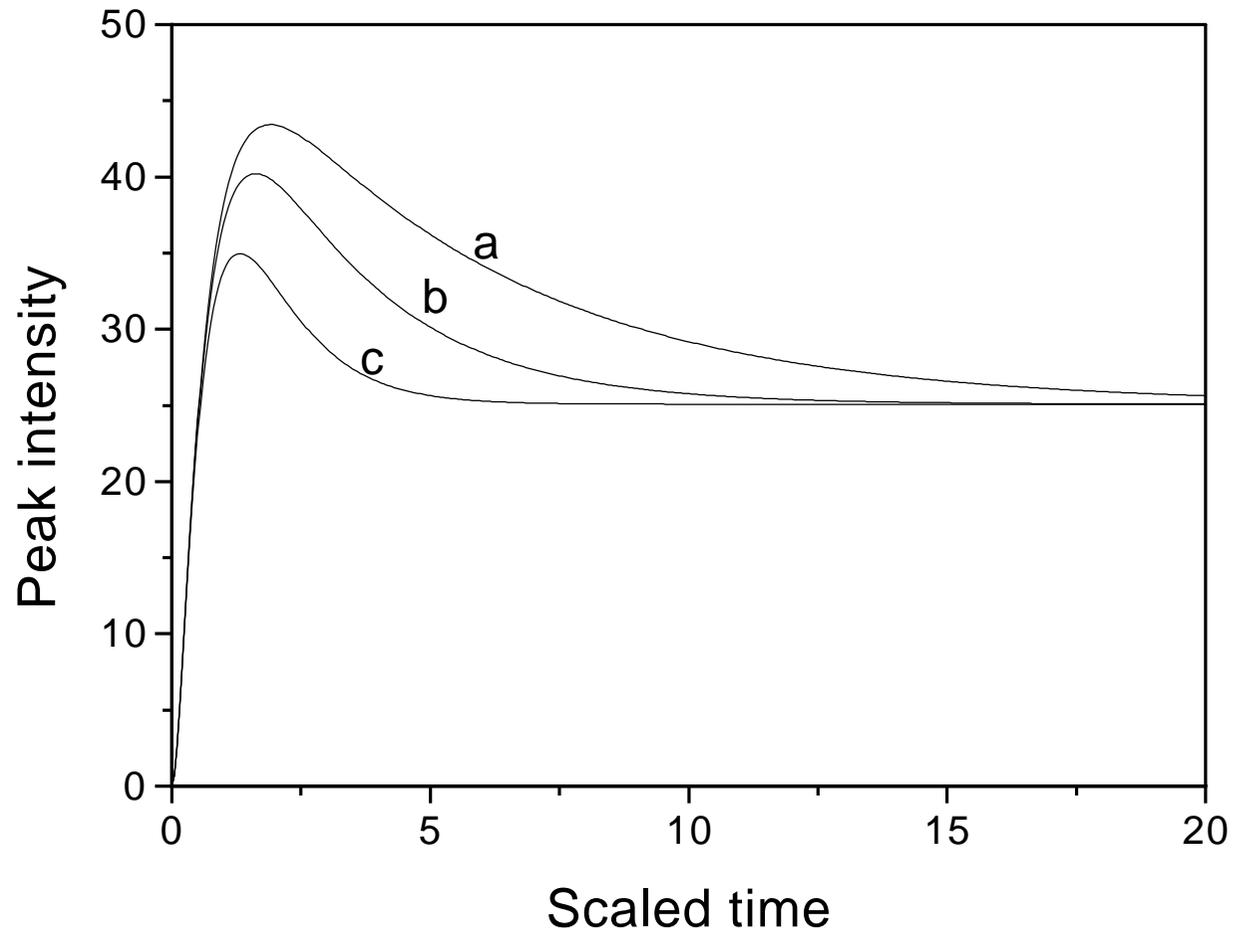
\end{document}